\documentclass[12pt,preprint]{aastex}

\shorttitle{{\em Spitzer} Observations of Cool WDs}
\shortauthors{Kilic et al.}

\begin{document}

\title{The First Mid-Infrared Spectra of Cool White Dwarfs}

\author{Mukremin Kilic\altaffilmark{1}, Piotr M. Kowalski\altaffilmark{2}, Fergal Mullally\altaffilmark{3},\\ William T. Reach\altaffilmark{4}, and Ted von Hippel\altaffilmark{5}}

\altaffiltext{1}{Department of Astronomy, Ohio State University, Columbus, OH 43210; kilic@astronomy.ohio-state.edu}
\altaffiltext{2}{Lehrstuhl f\"ur Theoretische Chemie, Ruhr-Universit\"at Bochum, 44780 Bochum, Germany}
\altaffiltext{3}{Department of Astrophysical Sciences, Princeton, NJ 08544}
\altaffiltext{4}{Spitzer Science Center, California Institute of Technology, Pasadena, CA 91125}
\altaffiltext{5}{Department of Astronomy, University of Texas, Austin, TX 78712}

\begin{abstract}

We present the first mid-infrared spectra of two cool white dwarfs obtained with the {\em Spitzer Space Telescope}.
We also present $3.5-8\mu$m photometry for 19 cool white dwarfs with 5000 K $\leq T_{\rm eff}\leq9000$ K.
We perform a detailed model atmosphere analysis of these white dwarfs by fitting their UBVRIJHK and {\em Spitzer} photometry with
state-of-the-art model atmospheres, and demonstrate that the optical and infrared spectral energy distributions of cool white dwarfs are well reproduced
by our grid of models. Our mid-infrared photometry and 7.5--14.5 $\mu$m spectrum of WD0018-267 are consistent with a $T_{\rm eff}=5720$ K, pure hydrogen
white dwarf model atmosphere. On the other hand, LHS 1126 remains peculiar with significant mid-infrared flux deficits in all
IRAC bands and a featureless spectrum in the 5.2--7.5 $\mu$m range. Even though this deficit is attributed to collision induced absorption (CIA)
due to molecular hydrogen, the shape of the deficit cannot be explained with current CIA opacity calculations. 
The infrared portion of the LHS 1126 spectral energy distribution is best-fit with a power law
index of $-$1.99; identical to a Rayleigh-Jeans spectrum. This argues that the deficit may be due to an unrecognized grey-like opacity source in the infrared.

\end{abstract}


\section{Introduction}

Cool white dwarfs (WDs) come in three different flavors: H-dominated, He-dominated, and mixed H/He atmospheres. The atmospheric
composition for WDs hotter than about 11,000 K can be easily identified from the observed hydrogen and helium lines.
However, helium lines disappear below 11000 K, and hydrogen becomes invisible below 5000 K. Optical spectroscopy cannot differentiate
between the three types of atmospheric compositions for WDs cooler than 5000 K. The dominant opacity sources in cool H-rich WD atmospheres are believed
to be CIA in the infrared (Hansen 1998; Saumon \& Jacobson 1999) and Ly$\alpha$ in the ultraviolet (Kowalski \& Saumon 2006; Koester \& Wolff 2000).
CIA opacity is strongly wavelength dependent and is expected
to produce molecular absorption features in the near-infrared (Frommhold 1993). For very cool WDs, these features extend into the optical
region (Harris et al. 2001; Gates et al. 2004).

The primary opacity source in He-rich WDs is He$^-$ free-free absorption (Kowalski et al 2007).
In mixed H/He atmospheres, collisions between H$_2$ molecules and neutral helium can induce CIA. Helium atmospheres are
much denser than the hydrogen atmospheres at the same temperature, therefore CIA becomes significant at higher temperatures compared
to pure hydrogen atmospheres. The only way to differentiate between pure H, pure He, and mixed H/He composition is to perform
a detailed model atmosphere analysis of both the optical and infrared colors of cool WDs. 

Cool WDs define the faint end of the WD luminosity function, and they are used to determine the ages of the Galactic
disk, halo, and nearby globular clusters (Winget et al. 1987; Liebert et al. 1988; Harris et al. 2006; Hansen et al. 2007).
Due to the different opacities, the atmospheric composition affects the WD cooling rate. Hence, determining
an accurate temperature, luminosity, and age for cool WDs requires determining the atmospheric compositions of the stars.
Bergeron et al. (2001) successfuly fit the optical and near-infrared colors of a sample of nearby WDs and they could explain the
cool WD SEDs by different atmospheric compositions. They found that the coolest WDs in the solar neighborhood have $T_{\rm eff}=$ 4200 K,
corresponding to an age of 8 Gyr. 

Kilic et al. (2006) presented {\em Spitzer} 4.5$\mu$m and 8$\mu$m photometry of 18 cool WDs, including 14 WDs from the Bergeron et al.
(2001) sample. Their preliminary analysis showed that several WDs with $T_{\rm eff} <$ 6000 K were slightly fainter than expected in the mid-infrared.
In addition, they found that LHS 1126, a WD believed to have mixed H/He atmosphere, displayed strong flux deficits in the 4.5$\mu$m
and 8$\mu$m bands. Fainter mid-infrared fluxes implied that the bolometric corrections for cool WDs may be wrong, and
the age estimates may be affected as well. Current CIA calculations do not predict absorption bands in the mid-infrared
which would imply that either the CIA calculations in WD atmospheres are incomplete, or these flux deficits are caused by some other
mechanism.

Tremblay \& Bergeron (2007) re-analyzed the {\em Spitzer} photometry for 12 of the WDs presented in Kilic et al. (2006), and with the
exception of LHS 1126, they did not find any significant flux deficit at low effective temperatures.
In order to understand the mid-infrared SEDs of cool WDs and the CIA opacity, we used the Infrared Spectrograph (IRS) on the
{\em Spitzer Space Telescope} to observe two nearby, relatively bright cool white dwarfs, WD0018-267 and LHS 1126\footnote{
Mid-infrared spectroscopy of two WDs (G29-38 and GD362) with $T_{\rm eff}= 1-1.2\times10^4$ K were presented by Reach et al. (2005a)
and Jura et al. (2007). However,
both of these stars have circumstellar debris disks, and their IRS spectra are dominated by the emission from the debris
disks. Hence, these spectra cannot be used to constrain the mid-infrared SEDs of cool WDs. In addition, our two targets are about 5000 K cooler than
GD362 which corresponds to an age difference of about 3 Gyr. Therefore, our IRS spectra of WD0018-267 and LHS 1126 are the first mid-infrared spectra of
cool WD photospheres.}.
In this paper, we present our mid-infrared spectroscopy of these two WDs and new IRAC photometry of several WDs. In order to resolve
the discrepancy between the Kilic et al. and Tremblay \& Bergeron analysis, we also re-analyze the photometry for 19 cool WDs.
Our IRAC and IRS observations are discussed in \S 2 and \S 3, respectively, while implications of these data are
discussed in \S 4. 

\section{IRAC Observations}

Observations reported here were obtained as part of the {\em Spitzer} Cycle 1 GO-Program 2313 (PI: M. Kuchner) and Cycle 2
GO-Program 20026 (PI: T. von Hippel). As part of our Cycle 1 program, we obtained 4.5 and 8$\mu$m data for 18 cool WDs
that were presented in Kilic et al. (2006). Here we re-analyze that data with the latest {\em Spitzer} Science Center
pipeline products.
As part of our Cycle 2 program, we used the Infrared Array Camera (IRAC; Fazio et al. 2004) to obtain 3.6 and 5.8$\mu$m
photometry of 4 cool white dwarfs including WD0018-267 and LHS1126.
An integration time of 30 seconds per dither, with five dithers for each target, was used.
The data reduction procedures are similar to the procedures employed by Kilic et al. (2006). The only difference in our reduction
procedures is the way we perform the array location dependent corrections for the IRAC instrument. Kilic et al. (2006) multiplied the BCD
images from the S11.1.0 pipeline with the location dependent correction image before performing photometry. We now perform aperture photometry
on the individual BCD frames from the S14.0.0 pipeline, and multiply the photometry by the coefficients obtained from the correction image supplied by the {\em Spitzer}
Science Center. This small difference in our procedure resulted in +2\% $\pm$ 0.7\% and $-$0.6\% $\pm$ 4\% differences in our 4.5$\mu$m and 8$\mu$m
photometry, respectively. The photometric error bars were estimated from the observed scatter in the 5 images (corresponding to 5 dither positions)
plus the 3\% absolute calibration error, added in quadrature.

Most of our objects were observed by Bergeron et al. (2001)
and therefore have accurate $BVRI$ photometry. For the remaining objects, we used the $UBV$ photometry from the McCook \& Sion (2006)
catalog. The near-infrared photometry comes from the 2MASS All-Sky Point Source Catalog (Cutri et al. 2003).
We fit the UBVRIJHK and {\em Spitzer} photometry for our targets with the current state of the art WD model atmospheres (including
the missing opacity from the Ly$\alpha$ wing; Kowalski \& Saumon 2006)
to determine $T_{\rm eff}$ and $\log$ g, if parallax measurements are available. Otherwise, we assume $\log$ g = 8. Our fitting
procedure is the same as the method employed by Tremblay \& Bergeron (2007).

The average fluxes measured from the {\em Spitzer} images are given in Table 1.
Figure 1 shows the optical and infrared SEDs (error bars) for 18 cool WDs with $T_{\rm eff}\leq$ 9000 K.
The expected fluxes from synthetic photometry of white dwarf model atmospheres
are shown as open circles.
This figure demonstrates that our grid of models reproduce the observed SEDs well in the entire range from 0.4$\mu$m to 8$\mu$m.
The agreement is better for the Bergeron et al. (2001) stars as the optical photometry is more accurate. We note that the U-band
photometry is slightly discrepant for some stars (e.g. WD0552-041 is a DZ, hence the U and B band photometry are affected by metals).
This is either caused by the uncertainties in the U-band observations obtained from the McCook \& Sion (2006) catalog,
or the presence of charged species in the atmosphere which may change the predicted U-band fluxes. Stark broadening
is not included in our Ly$\alpha$ model that is constructed for neutral medium. Stark broadening
becomes significant at $T_{\rm eff} \geq 7000$ K, and hence the U-band flux for stars hotter than 7000 K may be over-estimated in our models.

Figure 2 presents the ratio of observed to predicted 4.5$\mu$m and 8$\mu$m fluxes for hydrogen (crosses) and helium (filled circles)
dominated atmosphere WDs. Our photometric methods are consistent with those used for the IRAC
absolute calibration (Reach et al 2005b), and therefore they should be accurate to 3\%.
We note that the 8 $\mu$m photometry of WD1121+216 is affected by a nearby star, and therefore is not included in the bottom panel.
The observed 4.5$\mu$m fluxes are completely consistent with the WD model predictions, showing
that on general there are not atmospheric deviations of WDs in the
mid-infrared. The 8$\mu$m fluxes are on average fainter by 4\% compared to expectations. However, given the 3\% uncertainty in the IRAC
absolute flux calibration, we conclude that the IRAC photometry of cool WDs are consistent with WD model atmospheres; as a corollary,
WDs provide a calibration verification independent of the A and K stars that were included in the absolute calibration of IRAC.

Using the photometry from Kilic et al. (2006), Tremblay \& Bergeron (2007) found that the 12 WDs that they have analyzed were on average
fainter by 2\% in the 4.5$\mu$m band, whereas they could fit the 8$\mu$m photometry fairly well with their models. Our new photometry
is 2\% brighter in the 4.5$\mu$m band compared to Kilic et al. (2006). Therefore, our observations are consistent with the Tremblay \& Bergeron
(2007) analysis as well.

Other {\em Spitzer} WD surveys also targeted several cool WDs. The DAZ survey of Debes et al. (2007) included 4 stars with $T_{\rm eff}\sim$ 5300-8500 K. The ratios
of the observed fluxes to the predicted fluxes for these stars are 0.96 $\pm$ 0.04 and 0.98 $\pm$ 0.05 in the IRAC2 and IRAC4 bands, respectively.
The metal-rich WD survey of Farihi et al. (2007) included 7 stars with $T_{\rm eff}\sim$ 5200-8000 K that have mid-IR flux distributions consistent with
the models.

\section{IRS spectroscopy}

A better way to test the agreement between the WD models and the mid-infrared observations is to compare spectroscopy. 
As part of our Cycle 2 GO-Program 20026 (PI: T. von Hippel), we used the Infrared Spectrograph (IRS; Houck et al. 2004) Short-Low module in the first order
to obtain 7.4-14.5 $\mu$m spectroscopy of WD0018-267 with a resolving power of $\sim$90. These observations consisted 13 cycles of 4 min exposures and were
performed on 2006 July 1 21:12 UT. In addition, as part of our Cycle 3 GO-Program 30208 (PI: M. Kilic), we used the IRS Short-Low module in the second order to obtain
5.2-8.7 $\mu$m spectroscopy of LHS 1126. These observations consisted 52 cycles of 1 min exposures and were performed on 2007 August 2 02:48 UT. Both observations consisted
a series of exposures obtained at two positions along the slit.  
The IRS images at each nod position were filtered such that all
pixels deviating far from a 5-pixel wide median in the dispersion
direction were replaced by that median. Then the nods were
differenced and the spectra extracted using the {\em Spitzer}
IRS Custom Extractor (SPICE) in ``optimal'' extraction mode. For LHS
1126 the first-minus-second nod difference was used, as the second-minus-first
difference was contaminated over half the slit by stray light.

Figure 3 presents the weighted mean of the spectra for WD0018-267 from the two nods
(jagged line) compared to our best fit model atmosphere fit (solid line) to the VJHK and IRAC1-4 band
photometry (error bars). Since the distance to the WD is not known, we assumed $\log$ g = 8. This spectrum shows that
WD model atmospheres are able to explain the observed SED of a 5700 K pure hydrogen atmosphere WD fairly well.
We note that we did not use the IRS spectrum in our model fits. In addition, we did not normalize the spectrum to match the IRAC photometry or
the expected SED from the model atmospheres.

Figure 4 shows the ultraviolet, optical and infrared SED of LHS 1126 including our new IRS spectrum.
A remarkable feature of this SED is that the optical portion can be fit with a $\sim$5400 K WD model, whereas the infrared flux is deficient, and it gradually
decreases with no observable molecular features. Red and blue solid lines show the WD model atmospheres assuming atmospheric parameters derived
by Wolff et al. (2002) and Bergeron et al. (1997), respectively. 
These models include all non-ideal physics of dense helium introduced recently into modeling: Ly$\alpha$ red wing opacity (Kowalski \& Saumon 2006),
refraction (Kowalski \& Saumon 2004), improved description of ionization equilibrium (Kowalski et al. 2007) and the non-ideal dissociation
equilibrium of H$_2$ (Kowalski 2006). The models include CIA absorption and they demonstrate that CIA is expected to create wiggles and bumps in the spectrum
up to 3$\mu$m, and the SEDs of cool WDs are expected to return to normal in the mid-infrared.
Our {\em Spitzer} photometry and spectroscopy do not reveal any wiggles due to molecular absorption.
The dashed line in Figure 4 shows the best-fit power law to the infrared portion of the LHS 1126 SED from 1$\mu$m to 8$\mu$m.
The best fit power law has an index of $-$1.99, identical to a Rayleigh-Jeans tail spectrum. The best fit power laws for the other 18 stars in our sample have indices from
$-$1.80 for 9000 K WDs to $-$1.39 for 5500 K WDs; LHS 1126 has a unique SED.

\section{Discussion}

Brown dwarfs and massive planets are also expected to suffer from CIA (Borysow et al. 1997).
Cushing et al. (2006) and Roellig et al. (2004) obtained mid infrared spectroscopy of M, L, and T dwarfs; the observed spectra are
in good agreement with the model atmosphere calculations, with only a few minor deviations.
The only exception to this is 2MASS 0937+2931, a T6 dwarf. This object has very blue near-infrared colors and it shows flux suppression
in the K-band (Burgasser et al. 2002). {\em Spitzer} IRS observations
showed that 2MASS 0937+2931 also shows flux suppresssion up to 7.5 $\mu$m compared to typical T dwarfs (Cushing et al. 2006).
Burgasser et al. (2002) and Cushing et al. (2006) claimed that this flux suppression may be caused by high pressure (high surface gravity)
or low metallicity atmospheres. The CIA opacities are expected to be stronger in higher pressure environments, and therefore collision-induced
H$_2$ 1--0 dipole absorption centered at 2.4$\mu$m (Borysow 2002) could explain the K-band flux suppression.
However, the absorption up to 7.5$\mu$m is not predicted by the current CIA opacity calculations. Burgasser et al. (2007) suggest that
the observed spectral peculiarities in ``blue L dwarfs'' can be explained by the presence of thin and/or large-grained condensate clouds.

We find that the cool WD model atmospheres are able to explain the optical and infrared photometry and spectroscopy of all WDs in our sample
except LHS 1126, a cool WD only 10 pc away. LHS 1126 displays optical absorption features attributed to the C$_2$H molecule. 
Wickramasinghe et al. (1982) were the first ones to detect near-infrared flux deficit for LHS 1126. Bergeron et al. (1994; 1997) explained this deficit
as CIA by molecular hydrogen due to collisions with He and found the $\log$ (He/H) ratio to be $\sim$1.86. Analyzing the UV, optical, and near-IR data,
Wolff et al. (2002) found that such a high H abundance would lead to strong Ly$\alpha$ absorption, and the UV data is better fit with a
$\log$ (He/H) ratio of 5.51 (see Fig. 4). Recently, Kilic et al. (2006) discovered that LHS 1126 also suffers from significant flux deficits
at 4.5$\mu$m and 8$\mu$m, which are not compatible with the current CIA calculations. Now, we know that this flux deficit is present in all
IRAC bands. Our IRS spectrum is also featureless and it is consistent with a Rayleigh-Jeans tail spectrum.
Even though LHS 1126 looks like a cool WD with $T_{\rm eff} = 5400$ K in the optical, its infrared SED is reminiscent of a hotter WD.
Our models cannot explain the observed flux deficits in the {\em Spitzer} observations.

LHS 1126, like 2MASS 0937+2931, suffers from an unexplained opacity source up to 8$\mu$m. Both stars are expected to show CIA.
This implies that either the current CIA opacity calculations are inadequate or the mid-infrared flux deficits observed in these stars are caused
by a grey-like absorption process not included in our model atmosphere calculations.
The input physics used in WD model atmospheres is mostly based on the ideal gas approximation. The extreme conditions in WD
atmospheres require a new look at dense medium effects on the equation of state, opacities, and radiative transfer (see Kowalski 2006).

Unlike brown dwarfs, WD spectra do not suffer from absorption due to water, methane, and ammonia in the mid-infrared, therefore they are better
suited to find and constrain additional opacity sources like CIA. Near/mid-infrared photometry of white dwarfs cooler than 5000 K will be essential to test these ideas.
Such a study is currently underway with the {\em Spitzer Space Telescope} infrared camera.

\acknowledgements
PMK acknowledges partial financial support from Ruhr-Universit\"at in Bochum.
Support for this work was provided by NASA through
awards issued by JPL/Caltech to the University of Texas and the Ohio State University.
This work is based in part on observations made with the {\em Spitzer Space Telescope}, which is operated by the Jet Propulsion
Laboratory, California Institute of Technology under NASA contract 1407.
This publication makes use of data products from the Two Micron All Sky Survey, which is a joint project of the University of Massachusetts and the Infrared Processing and Analysis Center/California Institute of Technology, funded by the National Aeronautics and Space Administration and the National Science Foundation.

\clearpage
\begin{deluxetable}{lllcccc}
\tabletypesize{\footnotesize}
\tablecolumns{7}
\tablewidth{0pt}
\tablecaption{Infrared Photometry of Cool White Dwarfs}
\tablehead{
\colhead{Object}&
\colhead{Spectral Type}&
\colhead{$T_{\rm eff}$(K)}&
\colhead{$F_{3.6\mu m}$(mJy)}&
\colhead{$F_{4.5\mu m}$(mJy)}&
\colhead{$F_{5.8\mu m}$(mJy)}&
\colhead{$F_{8.0\mu m}$(mJy)}
} \startdata
WD0009+501  & DAP & 6631 & \nodata & 0.9562 $\pm$ 0.0514 & \nodata & 0.3159 $\pm$ 0.0186\\
WD0018$-$267& DA  & 5720 & 4.4218 $\pm$ 0.1396 & 2.8525 $\pm$ 0.0956 & 1.9086  $\pm$ 0.0699 & 1.0174 $\pm$ 0.0435\\ 
WD0038$-$226& C$_2$H:& 5400: & 0.7996 $\pm$ 0.0307 & 0.4943 $\pm$ 0.0162 & 0.3068 $\pm$ 0.0317 & 0.1953 $\pm$ 0.0319\\
WD0101+048  & DA  & 8360 & \nodata & 0.7693 $\pm$ 0.0264 & \nodata & 0.2808 $\pm$ 0.0307\\
WD0126+101  & DA  & 8926 & \nodata & 0.4592 $\pm$ 0.0214 & \nodata & 0.1364 $\pm$ 0.0668\\
WD0141$-$675& DA  & 6577 & 2.5736 $\pm$ 0.0810 & 1.6736 $\pm$ 0.0569 & 1.0423 $\pm$ 0.0503 & 0.5788 $\pm$ 0.0246\\ 
WD0552$-$041& DZ  & 5509 & \nodata & 1.8454 $\pm$ 0.0660 & \nodata & 0.7383 $\pm$ 0.0666\\ 
WD0553+053  & DAP & 5869 & \nodata & 1.6701 $\pm$ 0.0518 & \nodata & 0.5185 $\pm$ 0.0536\\ 
WD0752$-$676& DA  & 5661 & \nodata & 2.1580 $\pm$ 0.0770 & \nodata & 0.7710 $\pm$ 0.0379\\
WD0839$-$327& DA  & 8797 & \nodata & 4.0963 $\pm$ 0.1409 & \nodata & 1.4482 $\pm$ 0.0559\\
WD0912+536  & DCP & 7660 & \nodata & 1.1166 $\pm$ 0.0356 & \nodata & 0.4023 $\pm$ 0.0330\\
WD1055$-$072& DC  & 7740 & \nodata & 0.6567 $\pm$ 0.0249 & \nodata & 0.2512 $\pm$ 0.0356\\
WD1121+216  & DA  & 7328 & \nodata & 0.8229 $\pm$ 0.0302 & \nodata & 0.3294 $\pm$ 0.0185\\
WD1202-232  & DA  & 9115 & \nodata & 1.9858 $\pm$ 0.0692 & \nodata & 0.6835 $\pm$ 0.0358\\ 
WD1223$-$659& DA  & 7826 & \nodata & 0.9217 $\pm$ 0.0438 & \nodata & 0.3529 $\pm$ 0.0690\\
WD1756+827  & DA  & 7259 & \nodata & 0.7614 $\pm$ 0.0237 & \nodata & 0.2601 $\pm$ 0.0280\\
WD1953$-$011& DAP & 7894 & \nodata & 1.1540 $\pm$ 0.0425 & \nodata & 0.4395 $\pm$ 0.0263\\
WD2140+207  & DQ  & 9700 & 1.8462 $\pm$ 0.0632 & 1.2033 $\pm$ 0.0397 & 0.7877 $\pm$ 0.0482 & 0.3948 $\pm$ 0.0143\\
WD2359-434  & DA  & 8633 & \nodata & 1.7898 $\pm$ 0.0583 & \nodata & 0.6075 $\pm$ 0.0636\\
\enddata
\tablecomments{8 $\mu$m photometry of WD1121+216 is affected by a nearby star.}
\end{deluxetable}

\clearpage
\begin{figure}
\hspace{-0.7in}
\includegraphics[angle=-90,scale=.75]{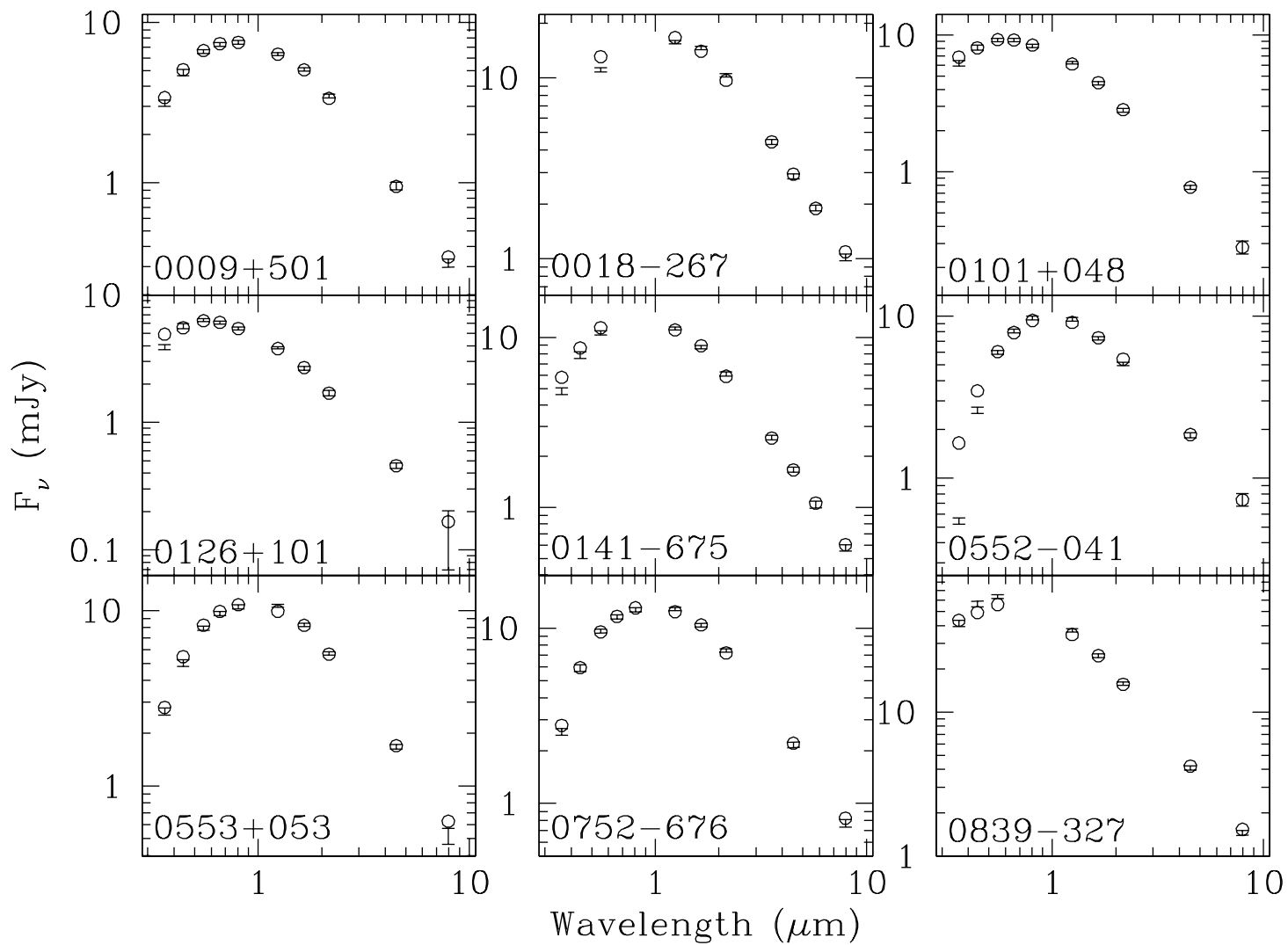}
\caption{Spectral energy distributions of cool white dwarfs observed in our Spitzer Cycle 1 and Cycle 2 programs. The observed
fluxes are shown as error bars, whereas the expected flux distributions from synthetic photometry
of white dwarf models are shown as open circles.}
\end{figure}
\clearpage
\centerline{\includegraphics[angle=-90,scale=.75]{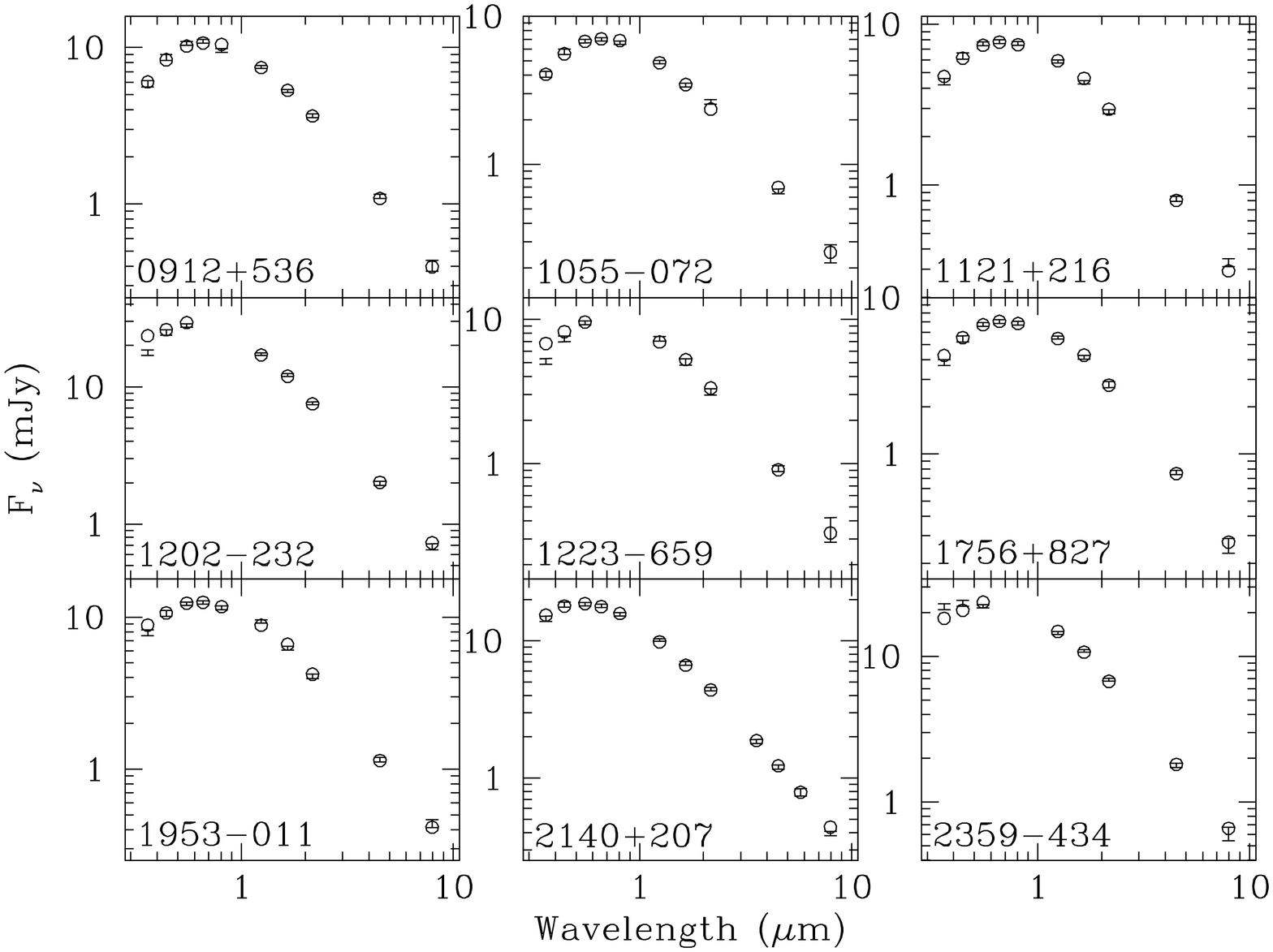}}
\centerline{Fig. 1. --- Continued.}

\clearpage
\begin{figure}
\includegraphics[angle=0,scale=.85]{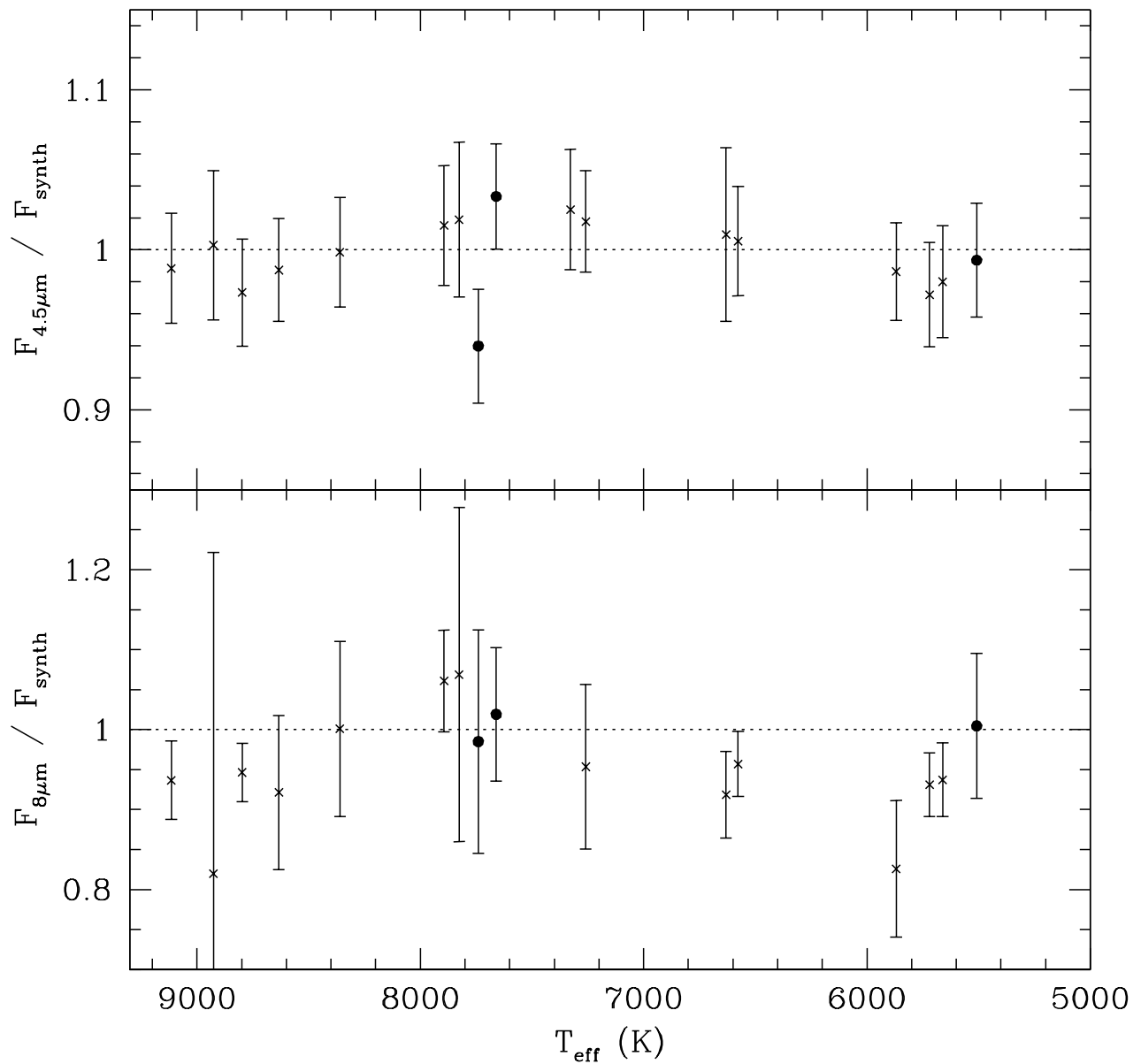}
\caption{Ratio of observed to predicted Spitzer fluxes for H (crosses) and He (filled circles) dominated atmosphere WDs.
The predicted fluxes and $T_{\rm eff}$ values are obtained from simultaneous fits to the UBVRIJHK and IRAC1-4 photometric data, where available.}
\end{figure}

\clearpage
\begin{figure}
\hspace{-0.7in}
\includegraphics[angle=-90,scale=.75]{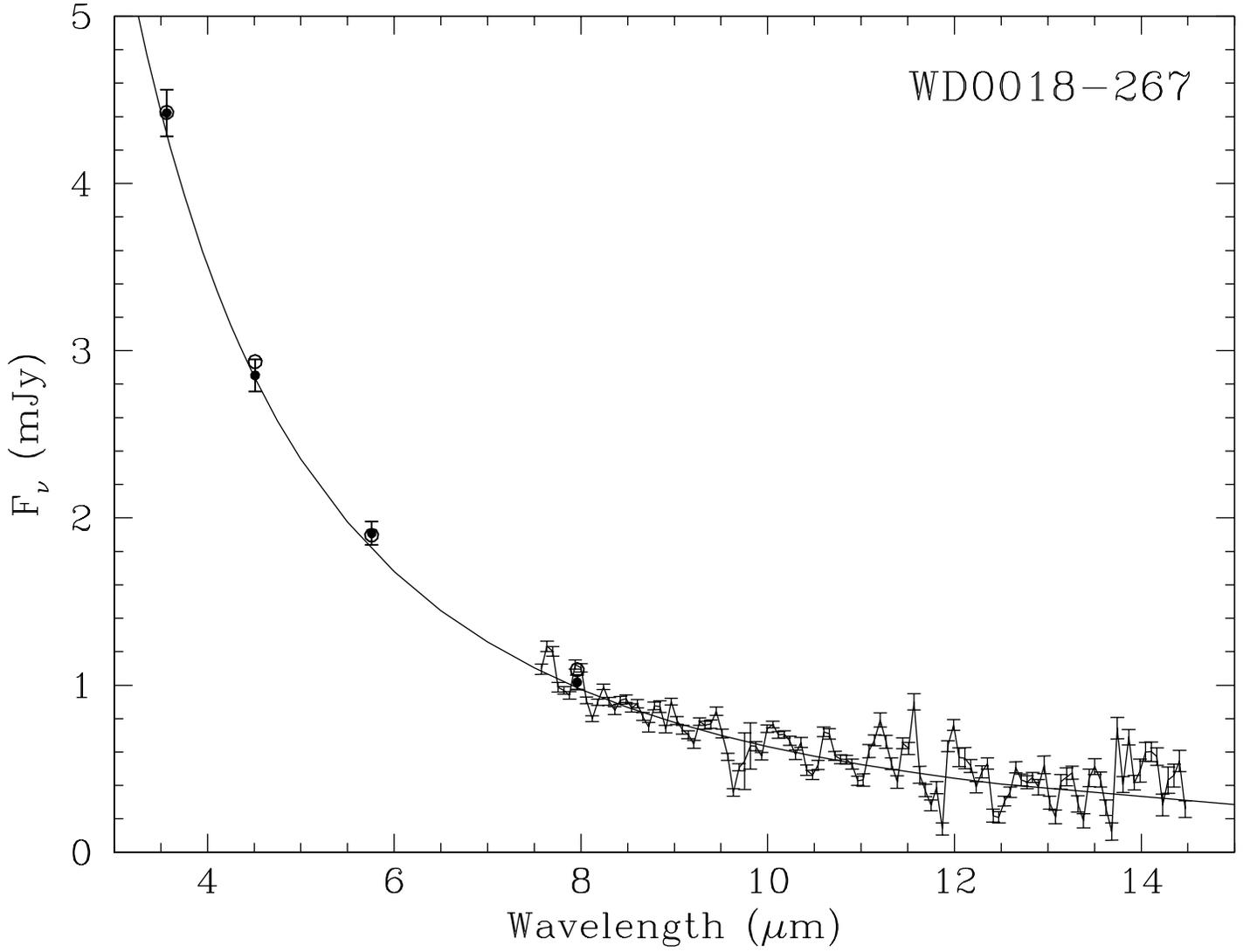}
\caption{IRAC photometry (filled circles with error bars) and IRS spectroscopy (error bars) of WD0018-267 compared to the best fitting pure H model atmosphere with
$T_{\rm eff}=$ 5720 K (solid line). Open circles show the synthetic photometry of the best fit model in the IRAC bands.}
\end{figure}

\clearpage
\begin{figure}
\hspace{-0.7in}
\includegraphics[angle=-90,scale=.75]{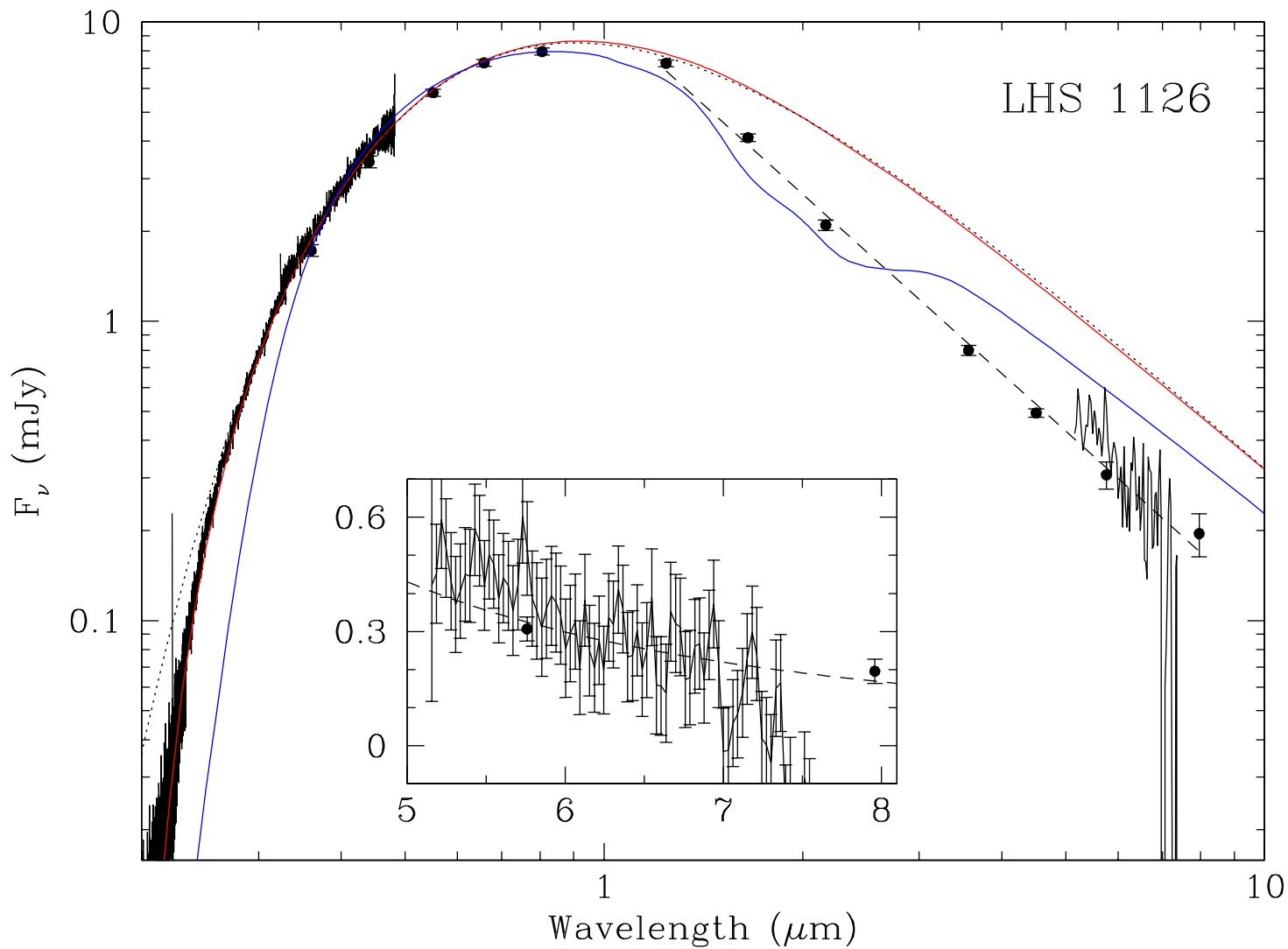}
\caption{Spectral energy distribution of LHS 1126, along with a 5400 K pure He atmosphere model (dotted line).
Red and blue solid lines show the model atmospheres using the best fit parameters from Wolff et al. (2002; $T_{\rm eff}=5400$ K, $\log$g = 7.9,
and $\log$ (He/H) = 5.51) and Bergeron et al. (1997; $T_{\rm eff}=5400$ K, $\log$g = 7.9, and $\log$ (He/H) = 1.86), respectively.
The dashed line shows a least square fit to the infrared photometry data only. The infrared portion of the spectral energy distribution is best fit
with a power law index of $-$1.99. The inset shows the 5-8$\mu$m spectral region in detail.}
\end{figure}

\end{document}